\documentclass[useAMS,usenatbib]{mn2e}

\usepackage{amssymb,amsmath,upgreek}

\usepackage{epsfig,graphicx,times,subfigure}

\newcommand{\msun}{M$_{\odot}$}

\newcommand{\lsun}{L$_{\odot}$}

\newcommand{\pasp}{PASP}
\newcommand{\apj}{ApJ}
\newcommand{\mnras}{MNRAS}
\newcommand{\araa}{ARA\&A}

\title[SN 2013ej]
{On the progenitor of the Type IIP SN 2013ej in M74}

\author[M. Fraser et al.]
{Morgan Fraser$^{1}$\thanks{E-mail:m.fraser@qub.ac.uk},
Justyn R. Maund$^{1}$\thanks{Royal Society Research Fellow},
Stephen J. Smartt$^{1}$,
Rubina Kotak$^{1}$,\newauthor
Andy Lawrence$^{2}$,
Alastair Bruce$^{2}$,
Stefano Valenti$^{3,4}$,
Fang Yuan$^{5}$,
Stefano Benetti$^{6}$\newauthor 
Ting-Wan Chen$^{1}$,
Avishay Gal-Yam$^{7}$,
Cosimo Inserra$^{1}$,
David R. Young$^{1}$\\ 
$^{1}$Astrophysics Research Centre, School of Mathematics and Physics, Queens University Belfast, Belfast BT7 1NN, UK\\
$^{2}$University of Edinburgh, Institute for Astronomy,  Royal Observatory, Blackford Hill, Edinburgh, EH9 3HJ, UK\\
$^{3}$Las Cumbres Observatory Global Telescope Network, 6740 Cortona Dr., Suite 102, Goleta, CA 93117, USA\\
$^{4}$Department of Physics, University of California, Santa Barbara, Broida Hall, Mail Code 9530, Santa Barbara, CA 93106-9530, USA\\
$^{5}$Research School of Astronomy and Astrophysics, Australian National University, Canberra, ACT 2611, Australia\\
$^{6}$INAF - Osservatorio Astronomico di Padova, vicolo dellÕOsservatorio 5, I-35122 Padova, Italy\\
$^{7}$Benoziyo Center for Astrophysics, Weizmann Institute of Science, Rehovot 76100, Israel
}

\begin{document}

\date{Submitted to Monthly Notices of the Royal Astronomical Society}

\pagerange{\pageref{firstpage}--\pageref{lastpage}} \pubyear{}

\maketitle

\label{firstpage}

\begin{abstract}
We use natural seeing imaging of SN 2013ej in M74 to identify a progenitor candidate in archival {\it Hubble Space Telescope} + ACS images. We find a source coincident with the SN in the {\it F814W}-filter, however the position of the progenitor candidate in contemporaneous {\it F435W} and {\it F555W}-filters is significantly offset. We conclude that the ``progenitor candidate'' is in fact two physically unrelated sources; a blue source which is likely unrelated to the SN, and a red source which we suggest exploded as SN 2013ej. Deep images with the same instrument onboard {\it HST} taken when the supernova has faded (in approximately two years time) will allow us to accurately characterise the unrelated neighbouring source and hence determine the intrinsic flux of the progenitor in three filters. We suggest that the {\it F814W} flux is dominated by the progenitor of SN 2013ej, and assuming a bolometric correction appropriate to an M-type supergiant, we estimate that the mass of the progenitor of SN 2013ej was between 8 -- 15.5 \msun.
\end{abstract}


\section{Introduction}
\label{s1}

Once a massive ($\gtrsim 8$ \msun) star has evolved through the stages of nuclear burning until it has an Fe core, it is no longer possible for it to generate enough energy to support the core against gravitational collapse. At this point, the star will explode as a core-collapse supernova (SN). Type II SNe result from the final explosion of a massive star which has retained its H envelope until the moment of collapse \citep{Fil97,Sma09a}. Type II SNe can be further classified as Type II Plateau (IIP) or Type II Linear (IIL) SNe, depending on whether they show a constant luminosity plateau or a linear decline in brightness between $\sim$30 and $\sim$100 days after explosion \citep{Bar79}. The diversity in Type II SN properties is thought to result chiefly from the degree to which they have retained a H-rich envelope \citep[e.g.][]{You04,Arc12}, and hence depends on progenitor mass, mass-loss, metallicity, binarity and rotation \citep{Lan12}.

Nearby ($\lesssim$30 Mpc) SNe are of interest not only for the detailed study they permit, but also because they raise the prospect of directly identifying their progenitors in pre-explosion images \citep[and references therein]{Van03a,Sma09b}. Red supergiant progenitors with masses between 8 -- 16 \msun\ have now been identified for around a dozen nearby Type IIP SNe \citep{Sma09b}. In a handful of cases, the progenitor candidate has been confirmed by its disappearance after the SN has faded \citep{Mau09}. 

The nearby galaxy Messier 74 (M74; also known as NGC 628) has hosted two previous SNe in the last two decades: the hydrogen poor Type Ic SN 2002ap, and the Type IIP SN 2003gd. In both cases, deep pre-explosion images were used to study the progenitor. For SN 2002ap, no source was identified in pre-explosion images down to very deep limits \citep{Sma02,Cro07}. In the case of SN 2003gd, both \cite{Van03b} and \cite{Sma04} found an 8-10 \msun\ red supergiant (RSG) coincident with the SN. \cite{Mau09} subsequently used late time imaging to show that this RSG was not longer present after the SN explosion.
 
The third supernova to be discovered in M 74 was found by the LOSS survey on 2013 July 25.5 UT, and designated SN 2013ej \citep{Kim13}. Spectroscopy from \cite{Bal13} and \cite{Val13a} confirmed that the object was a Type II SN discovered soon after explosion, and a preliminary progenitor identification was made by \cite{Van13}. In this Letter, we present an analysis of pre-explosion images of the site of SN 2013ej, and characterise the progenitor using extant archival data. A companion paper \citep{Val13b} presents the early photometric and spectroscopic coverage of SN 2013ej, showing that it is a bright Type IIP SN.

Despite the proximity of M74, it does not have a measured Cepheid or tip of the red giant branch distance. We have hence followed the approach of \cite{Hen05} and taken the mean of the distance to M74 derived from a range of techniques. In addition to the standard candle method distance, the brightest supergiants distance, and the kinematic distance used by \cite{Hen05}, we have included the \cite{Her08} planetary nebula luminosity function distance. The average of all four methods is 9.1$\pm$1.0 Mpc, where the error is given by the standard deviation among the measurements; we have used this distance in all of the following. We adopt a foreground reddening towards M74 of A$_V$=0.192 mag \citep{Sch11}.

\section{Archival data and progenitor identification}
\label{s2}

The {\it Hubble Space Telescope} observed the location of SN 2013ej in {\it UBVI}-like filters using the Wide Field and Planetary Camera 2 (WFPC2) and the Advanced Camera for Surveys (ACS), as detailed in Table \ref{tab:data}. To complement this data, we searched the publicly accessible archives of ground based 4m and 8m-class telescopes. All imaging which was of sufficient quality and depth to be of use is listed in Table \ref{tab:data}.

\begin{table*}
\caption{Log of observations for the candidate of progenitor SN 2013ej in pre-explosion images.}
\begin{center}
\begin{tabular}{lcccccccc}
\hline
Date			& Telescope 		& Instrument		& Filter		& Exposure (s)	& Resolution		& Magnitude	 	\\	
\hline
2003 Nov 20	& {\it HST}			& ACS			& {\it F435W}	& 8x590		& 0.05\arcsec		& 25.12 (0.06)	\\
2003 Nov 20	& {\it HST}			& ACS			& {\it F555W}	& 4x550		& 0.05\arcsec		& 24.84 (0.05)	\\
2003 Nov 20	& {\it HST}			& ACS			& {\it F814W}	& 4x390		& 0.05\arcsec		& 22.66 (0.03)	\\
2003 Dec 29	& {\it HST}			& ACS			& {\it F555W}	& 2x530		& 0.05\arcsec		& 25.01 (0.04)	\\
2005 Feb 16	& {\it HST}			& WFPC2			& {\it F336W}	& 4x1200		& 0.10\arcsec		& 23.31 (0.14) 	\\
2005 Jun 16	& {\it HST}			& ACS			& {\it F435W}	& 2x400		& 0.05\arcsec		& 25.16 (0.07)	\\
2005 Jun 16	& {\it HST}			& ACS			& {\it F555W}	& 1x360 		& 0.05\arcsec		& 25.16 (	0.09) 	\\
2005 Jun 16	& {\it HST}			& ACS			& {\it F814W}	& 2x360		& 0.05\arcsec		& 22.66 (0.03)	\\
2008 Sept 6	& Gemini N		& GMOS			& {\it r}'			& 1590		& 0.04\arcsec		& 23.89 (0.08)	\\
2008 Sept 6	& Gemini N		& GMOS			& {\it i}'			& 3180		& 0.04\arcsec		& 22.46 (0.09)	\\
\hline
\end{tabular}
\end{center}
\label{tab:data}
\end{table*}

To identify the position of SN 2013ej on the pre-explosion images, on 2013 August 8.2 UT we took a series of SDSS {\it r}-filter images using ACAM on the 4.2 m William Herschel Telescope, which provides an 8\arcmin\ field of view with 0.25\arcsec\ pixels. The brightness of SN 2013ej meant that saturation of the core pixels would occur in a few seconds, but these short exposures would not be deep enough for accurate alignment with the deep pre-explosion images. Hence, a set of short and long exposures were taken.  Frames of exposure times 1 sec, 30 sec and 3$\times$300 sec were taken while guiding smoothly during the sequence. 
The SN centroid was saturated in all images longer than 1 sec (0.85\arcsec\ FWHM image quality), but was not saturated in the shorter exposures (which had FWHM = 0.7\arcsec). 
The images were debiased and flatfielded using twilight flats and standard methods within {\sc iraf}\footnote{{\sc iraf} is distributed by the National Optical Astronomy Observatory, which is operated by the Association of Universities for Research in Astronomy (AURA) under cooperative agreement with the National Science Foundation.}. 14 stars with high significance detections (approximately greater than 10$\sigma$) where identified in common to the 1 sec and 30 sec frames and the short frame was aligned to the 30 sec frame (with pixel shifts of $-0.3$, 0.8 applied in $x,y$). The 3$\times$300s frames were combined into one and this frame also aligned to the 30 sec frame using 17 stars in common (pixel shifts of $-0.16, -0.63$ were found). In this way the 900s exposure was aligned to the 1s frame with an accuracy of $\pm0.02$ pixels in each dimension. The position of SN2013ej was then measured on the 1 sec frame using three different centring algorithms, and the measured value was assumed to be applicable to the 900s frame to this precision of $\pm0.02$ pixels. The total error on the SN position, estimated from the standard deviation of the three measurements of its position, plus the error in shift between the 1s and 900 s frame, was 18 mas.

The stacked 900s ACAM image was aligned to the drizzled, distortion-corrected 720s ACS/WFC {\it F814W} image of M 74 taken on 2005 June 16, which was obtained from the Hubble Legacy Archive (HLA)\footnote{hla.stsci.edu; filename: \texttt{HST\_10402\_22\_ACS\_WFC\_F814W\_drz.fits}}. Two separate alignments were made. In the first instance, we identified 28 point sources across both ACS chips, and measured their pixel coordinates in the ACS and the ACAM images. The matched coordinates were then used with {\sc iraf geomap} to derive a transformation between the two pixel coordinate systems. As there were a large number of reference sources for the alignment, we used a ``general'' fit within {\sc geomap}, which consists of a shift, scaling, rotation and a skew term. The residual of the fit was 95 mas. The measured position of the SN was then transformed to the pixel coordinates of the ACS frame. An obvious source was present at the transformed position, well within the total uncertainty (97 mas) in the alignment procedure. The entire procedure was then repeated, but using only sources within a 75\arcsec\ radius of SN 2013ej for the alignment. The reference sources for the second alignment were in general detected at a lower S/N, but have the advantage of being closer to the SN position and on the same chip.  33 sources were used, giving an rms error in the fit of 73 mas, and a total uncertainty in the SN position on the pre-explosion image of 75 mas. The same source (henceforth referred to as the progenitor candidate) was found to be coincident with the SN using both procedures. We measured the pixel coordinates of the progenitor candidate to be 3809.09, 2300.27 in the pre-explosion image, which is offset by 8 and 49 mas from our transformed positions, i.e. within the uncertainties.

\section{Progenitor analysis}
\label{s3}

As stated previously, the pre-explosion image on which the progenitor was identified was a pipeline drizzled {\it F814W} frame obtained from the HLA. The pixel scale of this image, 0.05\arcsec\ / pixel, is the native pixel scale of ACS. However, in the case of multiple observations taken with non-integer dithers, it is possible to reconstruct a combined image with a higher spatial resolution than the individual input frames, using the technique of drizzling \citep{Fru02}. Using the {\sc drizzlepac} package provided by the Space Telescope Science Institute\footnote{http://www.stsci.edu/hst/HST\_overview/drizzlepac}, we drizzled the ACS images for each filter taken in 2003 to a pixel scale of 0.03\arcsec / pixel. The pixel scale was chosen to provide the finest possible pixel scale, while minimising correlated noise and other artefacts introduced by the process of drizzling. We note that the drizzled frames do not permit a more accurate position for the progenitor to be determined, as the limiting factor in this case is the resolution of the post-explosion ACAM image.

\begin{figure}
\includegraphics[width=1\linewidth,angle=0]{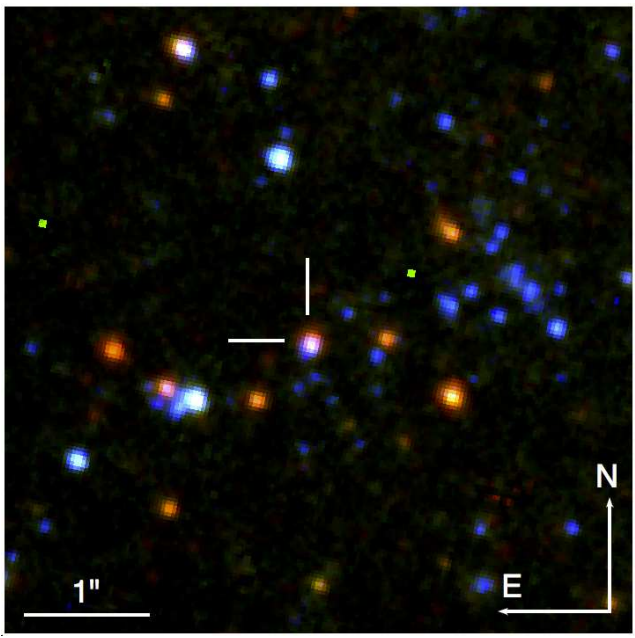}
\caption{{\it HST}+ACS {\it F435W/F555W/F814W} colour composite of the site of SN 2013ej. The SN position is indicated with cross marks; the offset between the source position in the blue and red filters is immediately apparent.}
\label{fig:progenitor}
\end{figure}

\begin{figure}
\centering
	\includegraphics[width=0.5\linewidth,angle=270]{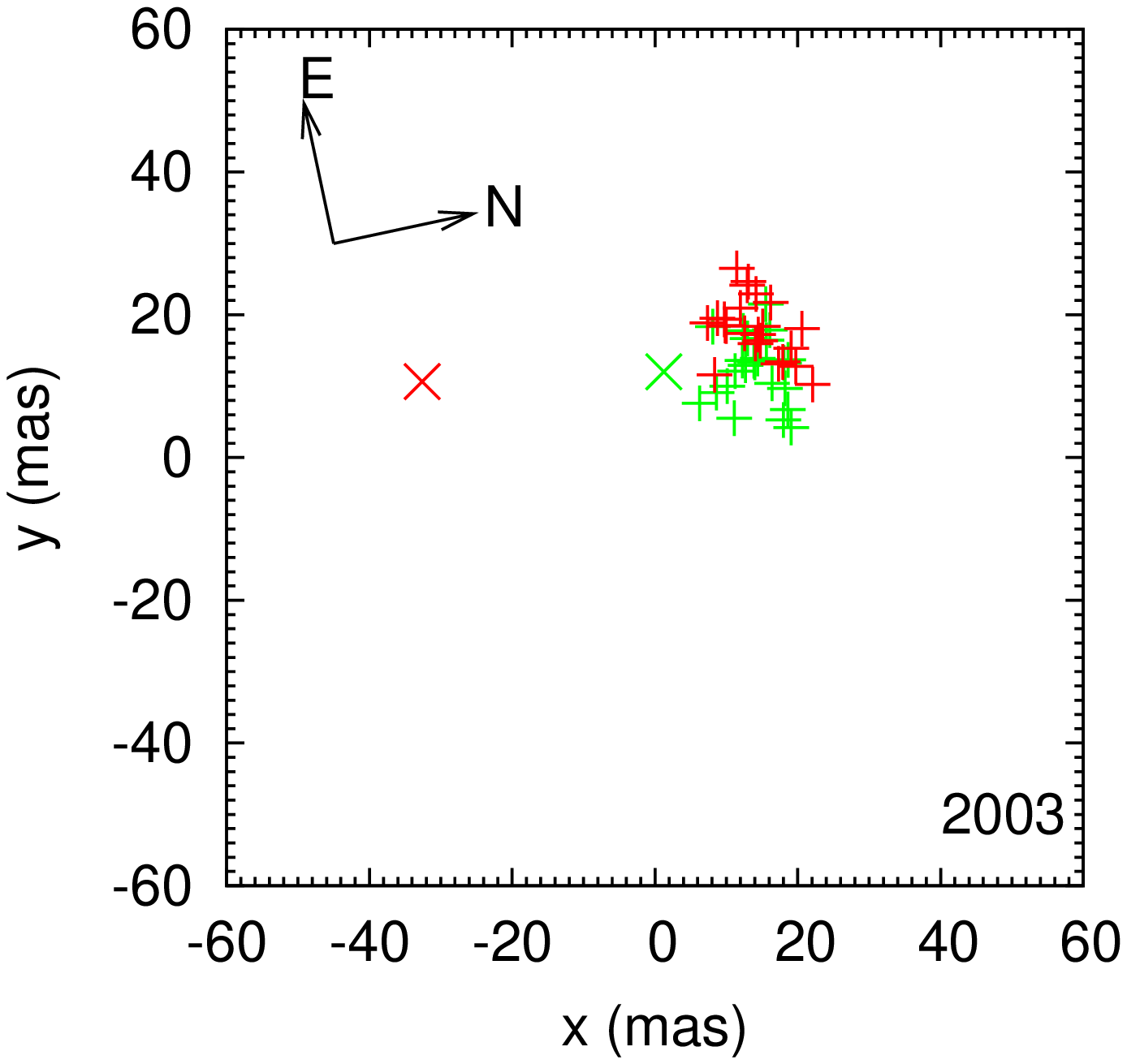}	
	\includegraphics[width=0.5\linewidth,angle=270]{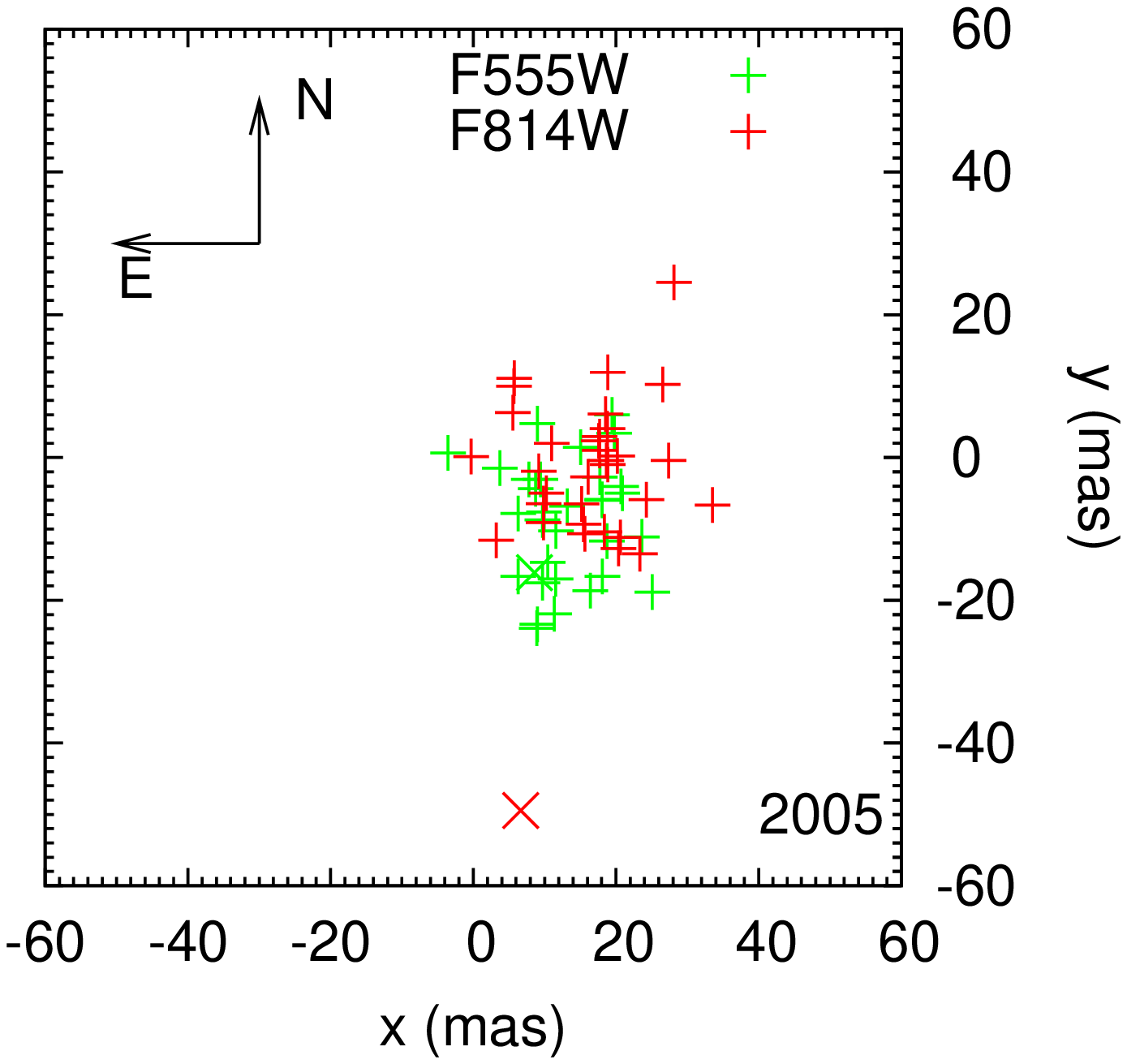}
\caption[]{Offset between the progenitor candidate position (source A+B) as measured in the {\it F555W} (green points) and {\it F814W} (red points) filter images, and the position measured in the {\it F435W} image taken at the same epoch. The progenitor candidate is indicated with an ``$\times$'', other sources in the vicinity which were used for a comparison are indicated with a ``$+$''.  The locus of points is offset slightly from 0,0 in each instance, this is the small sub pixel shift between the frames. There is an {\it additional} shift in the progenitor candidate position in the {\it F814W} image relative to its pixel coordinates as measured in the {\it F435W} image. This shift is consistent in direction and magnitude between the 2003 and 2005 data (note that the orientation of each panel is different). There is a smaller scatter in the 2003 data, which was drizzled to a finer pixel scale.}
\label{fig:offsets}
\end{figure}

We checked the position of the progenitor candidate in the various filter ACS images taken in both 2003 and 2005. For the former, we used the drizzled images, while for the 2005 data we were unable to improve on the spatial resolution by drizzling, and so used the HLA images at the native 0.05\arcsec\ pixel scale. We combined the drizzled {\it HST}+ACS {\it F435W}, {\it F455W} and {\it F814W} images taken on 20 November 2003, to an accuracy of 4 mas to create a colour composite of the site of SN 2013ej, as shown in Fig. \ref{fig:progenitor}. It is immediately evident that the blue and red flux from the progenitor are spatially offset, suggesting that this is neither a single star nor a compact stellar cluster. In order to quantify this further, in both the 2003 and 2005 data we measured the pixel coordinates of 20-30 point-like reference sources which were visible in all filters, together with the position of the progenitor candidate. The resulting offsets are shown in Fig. \ref{fig:offsets}. For both epochs we find that the position of the progenitor candidate differs by $\sim$40 mas in the {\it F814W} and {\it F435W} filters. In the 2003 data, we measure a difference of 47 mas between the progenitor candidate offset (from {\it F435W} to {\it F814W}) and the mean offset of the reference sources. The standard deviation of the sample of reference source offsets is only 6 mas, hence 47 mas is a statistically significant 8$\sigma$ difference. We see no significant offset for the position in {\it F555W} with respect to {\it F435W}.

We interpret this offset between the progenitor candidate position in the {\it F814W} image and the position in the {\it F435W} and {\it F555W} images, as resulting from two separate, physically unrelated sources. One source  (henceforth ``Source A'') dominates the flux in {\it F814W}, while a second source (``Source B'') contributes most of the flux in {\it F435W} and {\it F555W}). The offset between Source A and Source B (47 mas) corresponds to 2 pc at the distance of M74, so it is not feasible for this to be a binary system. In the remainder of this section we perform photometry on the combined Source A + Source B, while in Section \ref{s4} we discuss the implications for the progenitor of SN 2013ej.

Photometry of the progenitor candidate (Source A+B) in the ACS images was performed on the original \texttt{\_flc} images using {\sc dolphot}, a photometry package adapted from {\sc hstphot} \citep{Dol00a}. The data were downloaded from the Mikulski Archive for Space Telescopes (MAST), and have been automatically reduced by the CALACS pipeline. The \texttt{\_flc} files have been corrected for charge-transfer efficiency (CTE) by reconstructing the flux in affected pixels, and so no CTE correction was applied to the measured magnitudes. The progenitor candidate (Source A+B) was clearly detected by {\sc dolphot} at a significance of between 20$<\sigma<$50 in all filters. The counts measured for the progenitor candidate were then converted to a magnitude in the VEGAMAG system by applying the most up to date zeropoint from the STScI webpages for the relevant epoch\footnote{http://www.stsci.edu/hst/acs/analysis/zeropoints}. The measured magnitudes and associated uncertainties are reported in Table \ref{tab:data}. We note that when using {\sc dolphot} in its default mode (i.e. applying the built-in zeropoint corrections), we recover magnitudes which are 0.06-0.08 mag fainter than \cite{Van13}. While this is outside our formal error, we regard this level of agreement as acceptable, given the slightly different results which are obtained from {\sc dolphot} depending on the precise choice of aperture and sky annulus.

\begin{figure*}
\includegraphics[width=0.25\linewidth,angle=270]{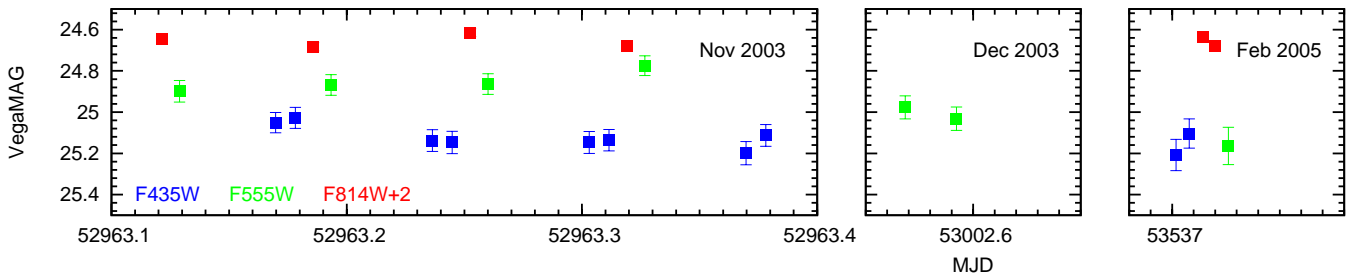}
\caption{{\it HST}+ACS photometry of the source coincident with SN 2013ej. In all panels, the tics on the x-axis correspond to 0.01 day intervals. The {\it F814W} filter magnitudes have a constant offset of +2 added to them.}
\label{fig:acsphot}
\end{figure*}

Using the average of the 2003 and 2005 ACS magnitudes, and correcting for foreground extinction, we find a {\it F435W-F555W} colour of 0.12 mag, and an {\it F555W-F814W} colour of 2.20 mag. While the latter is consistent with a RSG progenitor as would be expected for a Type IIP SN, the former is too blue for an RSG. This apparent inconsistency is further evidence that two objects are contributing to the measured flux.  We plot a lightcurve using the ACS photometry for Source A+B in Fig. \ref{fig:acsphot}. We see evidence for some variability in  the {\it F555W} filter, but as this is dominated by Source B, this is unlikely to be connected to the progenitor. In the {\it F814W} filter (which we assume is largely due to Source A) we see no evidence for variability above the 0.05 mag level.

WFPC2 observed the site of SN 2013ej in the {\it F336W} filter. Photometry was performed on these images using {\sc hstphot} \citep{Dol00a}. A 4-6$\sigma$ source was detected by {\sc hstphot} at the position of SN 2013ej in three of the four individual {\it F336W} filter exposures, giving a combined significance for the detection of 8$\sigma$. The {\it F336W} magnitude of the source coincident with SN 2013ej is given in Table \ref{tab:data}, however given that we associated the {\it F435W} and {\it F555W}-band flux with Source B, it seems likely that the {\it F336W}-band flux is also unrelated to the progenitor.

The site of SN~2013ej was also observed prior to explosion with the 
Gemini GMOS-N, in September 2008 (conducted as part of program 
GN-2008B-Q-67; PI Maund). The observations were conducted under 
excellent seeing ($0.35\arcsec$  in $i'$) and photometric conditions. 
The reduction and analysis of this data has been previously presented by 
\cite{Mau09}.  In the $i'$-band, we find a source coincident 
with the SN position.  The results of PSF photometry of the progenitor 
candidate in the GMOS images are also presented in Table \ref{tab:data}.    The $i'$ 
photometry of the pre-explosion source is $\sim$ 0.2 mag brighter than 
the corresponding ACS {\it F814W} photometry (not corrected for 
differences between the filter transmission functions).  Despite the 
high quality of these ground-based images, we cannot resolve the two 
sources observed at the SN position in the {\it HST} images; and the 
pre-explosion source is partially blended with a number of objects in 
close proximity.  We note that the properties of the PSF fit for the 
pre-explosion source are relatively poor ($\chi^2_{red}=2.9$ in $i'$); and in 
$g'$ we cannot identify a single source exactly coincident with the SN 
position with confidence.
The brighter photometry measured from the Gemini GMOS $i'$ image may 
reflect blending with nearby sources, incorrectly accounted for in the 
PSF fit.  Given these uncertainties, we can only note that the $i'$ 
photometry is not significantly discrepant from the {\it HST} photometry so 
as to indicate large variability at the SN position prior to explosion.

M74 has been observed by the {\it Spitzer Space Telescope} + IRAC in Ch1 and Ch2 (3.6$\upmu$m and 4.5$\upmu$m respectively). The resolution of IRAC is $\sim$1.7\arcsec\ with 1.2\arcsec\ pixels, hence the camera does not have the spatial resolution necessary to resolve a single stellar progenitor at this distance. We examined the 3.6 $\upmu$m image analysed by \cite{Kha13}, and see flux at the progenitor position, however from comparison to the {\it HST} images it is clear that this flux comes from a blend of multiple sources. While we have not considered the IRAC images any further here, they may be of use in the future with template subtraction, when deep images without the progenitor can be obtained after SN 2013ej has faded.

\section{Discussion}
\label{s4}

Once SN 2013ej has faded below the magnitude of the progenitor candidate, it will be relatively straightforward to obtain deep imaging of M74, and perform image subtraction to determine any decrease in flux since 2003 due to the disappearance of the progenitor. Such an approach has already been used successfully for other Type IIP SNe \citep{Mau13}, and for the Type IIn SN 2005gl \citep{Gal09}. Until then, we can estimate a progenitor mass from the {\it F814W} magnitude. Assuming all the flux in this filter comes from the progenitor, and that the progenitor was a RSG with a temperature between 3400-4000 K (appropriate for a late K to M-type supergiant), we can derive a luminosity. The assumption that the progenitor is an RSG is reasonable, both given RSG progenitors seen for other Type IIP SNe, and the spectrophotometric evolution of SN 2013ej which \cite{Val13b} have shown is consistent with an extended H-rich progenitor. We take bolometric corrections and colours from MARCS stellar atmosphere models \citep{Gus08}. Using these, we find a range of progenitor luminosities between log L/\lsun = 4.46--4.85 dex, depending on the distance and bolometric correction applied.

Using the STARS code, this luminosity corresponds to the final (strictly, at the beginning of core Ne burning) luminosity of a SN progenitor in the mass range 8-15.5\msun. Similar to \cite{Sma09b}, we set an upper limit to the progenitor mass by comparing the maximum luminosity of the progenitor candidate to that of the luminosity of models at the end of core He burning. The luminosity at the end of He burning is the {\it minimum} luminosity a star could have at the point of core-collapse, and so this is a conservative approach to deriving a maximum progenitor mass. The upper mass limit is also conservative to any contribution of flux in the {\it F814W} filter from Source B, as this will only lead to an over-estimate of the progenitor luminosity, and hence mass.

One final caveat which must be applied to our result is that circumstellar extinction around the progenitor could cause the {\it F814W} flux to be underestimated. The spectra and photometry of SN 2013ej do not appear to be significantly reddened \citep{Val13b}. However, in the case of the Type IIP SN 2012aw \citep{Fra12,Van12}, significant pre-existing circumstellar dust was believed to be destroyed in the SN explosion, resulting in a relatively high progenitor mass estimate, although \cite{Koc12} subsequently revised this estimate downwards. In the absence of multi-colour imaging and a measured colour for the progenitor candidate, this effect is impossible to quantify, although  theoretical models suggest that the effect of intrinsic circumstellar dust on a progenitor mass estimate derived from {\it I}-band photometry should be $<$1 \msun\ \citep{Wal12}. Furthermore, high resolution spectroscopy of SN 2013ej shows no strong Na{\sc i}~D absorption \citep{Val13b}, although this does not preclude dust destroyed by the shock breakout of the SN.

The proximity of SN 2013ej presents a relatively rare opportunity to intensively follow a Type IIP SN until very late phases. It is hence of great value to know what type and mass of progenitor exploded. While the spectral type and temperature of the progenitor remains unknown, we have presented a compelling argument that the progenitor mass is likely $<$16 \msun. As such, SN 2013ej joins an ever growing population of Type IIP SNe which appear to come from 8--16 \msun\ progenitors, and provides further evidence for a surprising absence of SNe resulting from high mass ($>$16 \msun) progenitors \citep{Koc08,Sma09b,Eld13}.

\section{Acknowledgements}

The research leading to these results has received funding from the
European Research Council under the European Union's Seventh Framework
Programme (FP7/2007-2013)/ERC Grant agreement n$^{\rm o}$ [291222] (PI
: S. J. Smartt).
The research of J.R.M. is funded through a Royal Society University Research Fellowship.
S.B. is partially supported by the PRIN-INAF 2011 with the project ``Transient Universe: from ESO Large to PESSTO''.
A.G. acknowledges support by the EU/FP7 via ERC grant n$^{\rm o}$ 307260, a GIF grant, and the
Kimmel award.

Partially based on observations made with the NASA/ESA Hubble Space Telescope, obtained from the data archive at the Space Telescope Science Institute. STScI is operated by the Association of Universities for Research in Astronomy, Inc. under NASA contract NAS 5-26555. Partially based on observations obtained at the Gemini Observatory, which is operated by the     Association of Universities for Research in Astronomy, Inc., under a cooperative agreement  with the NSF on behalf of the Gemini partnership: the National Science Foundation  (United States), the National Research Council (Canada), CONICYT (Chile), the Australian     Research Council (Australia), Minist\'{e}rio da Ci\^{e}ncia, Tecnologia e Inova\c{c}\~{a}o  (Brazil) and Ministerio de Ciencia, Tecnolog\'{i}a e Innovaci\'{o}n Productiva (Argentina). This research has made use of the NASA/IPAC Extragalactic Database (NED) which is operated by the Jet Propulsion Laboratory, California Institute of Technology, under contract with the National Aeronautics and Space Administration.

M.F. thanks John Eldridge for suggestions and advice.

\bsp

\label{lastpage}

\end{document}